\documentclass{appolb}

\usepackage[final]{graphics}
\usepackage{bbm}
\usepackage{amsmath}
\usepackage{amsfonts,amsbsy}
\usepackage{amssymb}
\usepackage{feynman}

\def\empile#1\over#2{\mathrel{\mathop{\kern 0pt#1}\limits_{#2}}}

\newcommand{\slv}{\raise.15ex\hbox{$/$}\kern-.53em\hbox{$v$}}
\newcommand{\slF}{\raise.15ex\hbox{$/$}\kern-.53em\hbox{$F$}}
\newcommand{\slL}{\raise.15ex\hbox{$/$}\kern-.53em\hbox{$L$}}
\newcommand{\slP}{\raise.15ex\hbox{$/$}\kern-.53em\hbox{$P$}}
\newcommand{\slp}{\raise.15ex\hbox{$/$}\kern-.53em\hbox{$p$}}
\newcommand{\slq}{\raise.15ex\hbox{$/$}\kern-.53em\hbox{$q$}}
\newcommand{\slR}{\raise.15ex\hbox{$/$}\kern-.53em\hbox{$R$}}
\newcommand{\slQ}{\raise.15ex\hbox{$/$}\kern-.53em\hbox{$Q$}}
\newcommand{\slK}{\raise.15ex\hbox{$/$}\kern-.53em\hbox{$K$}}
\newcommand{\slk}{\raise.15ex\hbox{$/$}\kern-.53em\hbox{$k$}}
\newcommand{\slD}{\raise.15ex\hbox{$/$}\kern-.53em\hbox{$D$}}
\newcommand{\slC}{\raise.15ex\hbox{$/$}\kern-.53em\hbox{$C$}}
\newcommand{\slA}{\raise.15ex\hbox{$/$}\kern-.53em\hbox{$A$}}
\newcommand{\slSigma}{\raise.15ex\hbox{$/$}\kern-.53em\hbox{$\Sigma$}}
\newcommand{\slpartial}{\raise.15ex\hbox{$/$}\kern-.53em\hbox{$\partial$}}
\newcommand{\slcalP}{\raise.15ex\hbox{$/$}\kern-.63em\hbox{$\cal P$}}
\newcommand{\slcalA}{\raise.15ex\hbox{$/$}\kern-.63em\hbox{$\cal A$}}

\def\x{{\boldsymbol x}}

\catcode`\@=11

\newcount\@tempcntc
\def\@citex[#1]#2{\if@filesw\immediate\write\@auxout{\string\citation{#2}}\fi
  \@tempcnta\z@\@tempcntb\m@ne\def\@citea{}\@cite{%
        \@for\@citeb:=#2\do%
    {\@ifundefined{b@\@citeb}%
        {\@citeo\@tempcntb\m@ne\@citea%
                \def\@citea{,\penalty\@m\ }{\bf ?}\@warning%
                {Citation `\@citeb' on page \thepage \space undefined}}%
        {\setbox\z@\hbox{\global\@tempcntc0\csname b@\@citeb\endcsname\relax}
     \ifnum\@tempcntc=\z@ \@citeo\@tempcntb\m@ne%
       \@citea\def\@citea{,\penalty\@m}%
       \hbox{\csname b@\@citeb\endcsname}%
     \else%
      \advance\@tempcntb\@ne%
      \ifnum\@tempcntb=\@tempcntc%
      \else\advance\@tempcntb\m@ne\@citeo%
      \@tempcnta\@tempcntc\@tempcntb\@tempcntc\fi\fi}}\@citeo}{#1}}%

\def\@citeo{\ifnum\@tempcnta>\@tempcntb\else\@citea
  \def\@citea{,\penalty\@m}%
  \ifnum\@tempcnta=\@tempcntb\the\@tempcnta\else
   {\advance\@tempcnta\@ne\ifnum\@tempcnta=\@tempcntb \else
\def\@citea{--}\fi
    \advance\@tempcnta\m@ne\the\@tempcnta\@citea\the\@tempcntb}\fi\fi}

\catcode`\@=12

\begin{document}
\title{Some aspects of ultra-relativistic heavy ion collisions
\thanks{Presented at the XXVIIth Physics In Collision conference, LAPP, Annecy, France.}%
}
\author{Fran\c cois Gelis
\address{CERN, PH-TH, CH-1211 Geneva 23, Switzerland}
}
\maketitle
\begin{abstract}
In this talk, I discuss some recent results obtained in Heavy Ion
Collisions and what they tell us -- or what questions they raise --
about the physics of the system of quarks and gluons formed in these
collisions.
\end{abstract}
  
\section{Quark-Gluon plasma and heavy ion collisions}
\begin{figure}[htbp]
\centerline{
\hfill
\resizebox*{5cm}{!}{\includegraphics{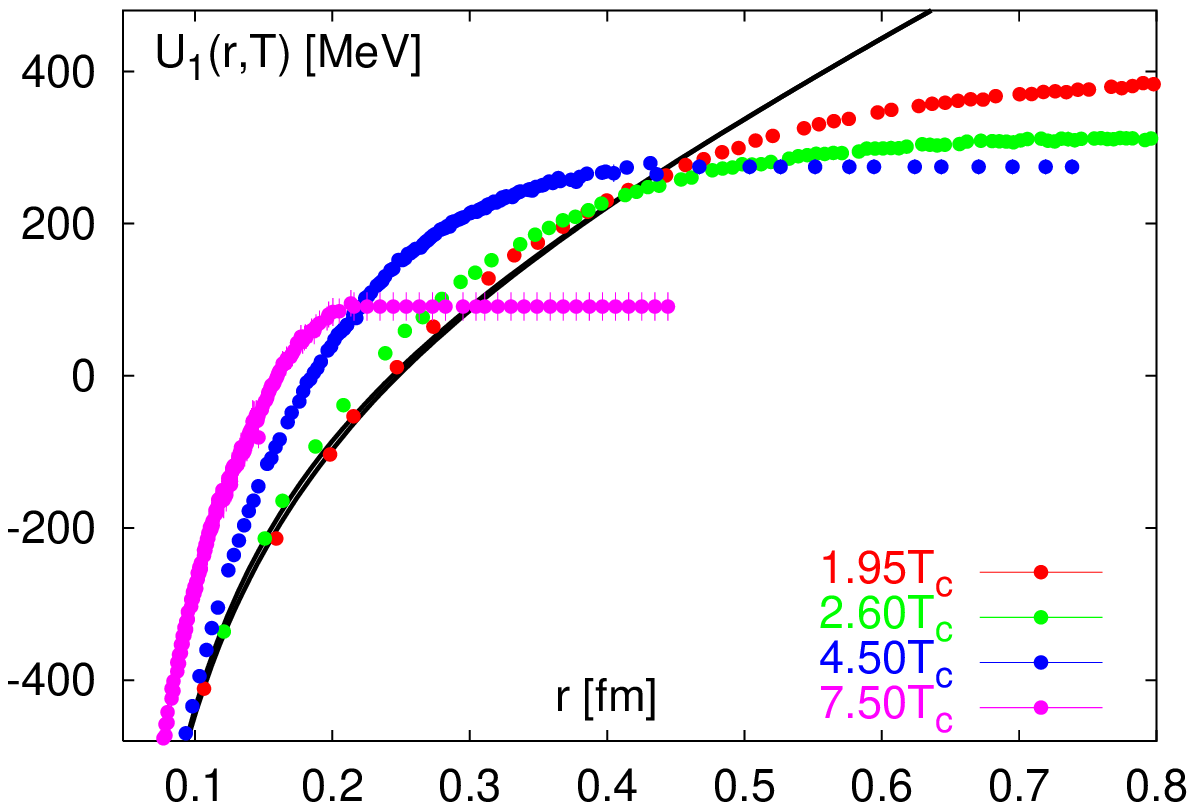}}
\hfill
\resizebox*{5cm}{!}{\includegraphics{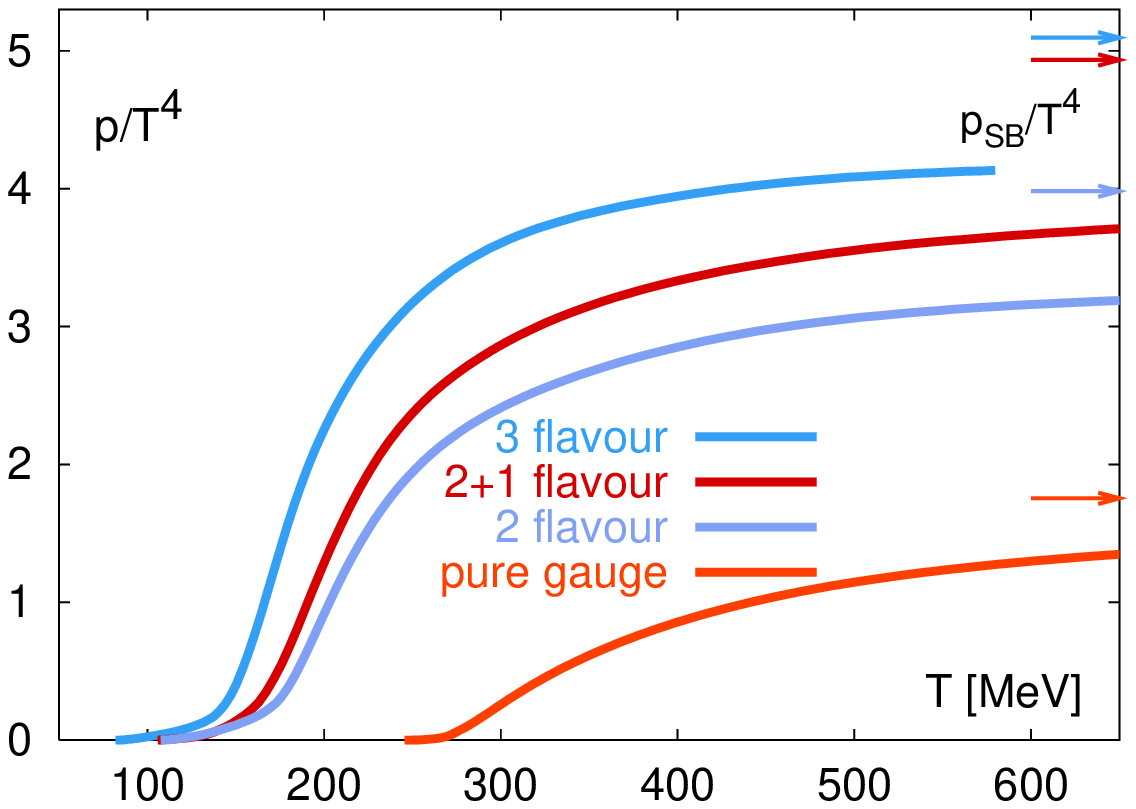}}
\hfill
}
\caption{\label{fig:transition}Lattice results. Left: quark potential
as a function of distance at various temperatures. Right: pressure as
a function of temperature.}
\end{figure}

The existence of a deconfined phase of nuclear matter was conjectured
long ago on the basis of asymptotic freedom, and has now received
ample support from QCD simulations on the lattice \cite{latt}. Some of
these results are displayed in figure \ref{fig:transition}.
On the left figure is displayed the potential between two static
quarks as a function of the distance between them. The solid line is
the zero temperature potential, that shows a linear rise at large
distance - a sign of quark confinement. The dotted lines show the same
potential at increasing temperatures. The main feature is that, while
the short distance behavior is not much affected, the linearly
increasing tail eventually disappears. This indicates that above a
certain temperature, it costs only a finite energy to separate the two
quarks. Another evidence for a phase transition is shown in the right
plot of figure \ref{fig:transition}, where one can see that the
pressure rises very rapidly at a certain temperature (the value of
which depends on the quark content of the theory), indicating a sudden
increase in the number of degrees of freedom in the system. This is
interpreted as a transition from hadronic bound states to a plasma of
quarks and gluons.

\setbox1\hbox to 10cm{
\hfil
\begin{feynman}{7.5cm}
\diagram{25}{
/recule {stringwidth neg exch neg exch rmoveto} def
-2 0.5 translate
fgcolor
0.08 setlinewidth
/LL 0.85 def
newpath
initclip
0 0 moveto 
0 0.5 0.6 1 LL 1.5 curveto 2 5 rlineto currentpoint
pop 2 mul neg 0 rlineto
2 -5 rlineto -0.6 1 0 0.5 0 0 curveto clip
newpath
0.8441 0.0000 0.0000 setrgbcolor
-3 0 moveto 6 0 rlineto 0 1.3 rlineto -6 0 rlineto 0 -1.3 rlineto fill
newpath
1.0000 0.2455 0.0000 setrgbcolor
-3 1.3 moveto 6 0 rlineto 0 1 rlineto -6 0 rlineto 0 -1 rlineto fill
newpath
1.00 0.75 0.00 setrgbcolor
-3 2.3 moveto 6 0 rlineto 0 1.3 rlineto -6 0 rlineto 0 -1.3 rlineto fill
newpath
0.8664 0.7223 0.5664 setrgbcolor
-3 3.6 moveto 6 0 rlineto 0 1.2 rlineto -6 0 rlineto 0 -1.2 rlineto fill
newpath
0.9664 0.8223 0.666 setrgbcolor
-3 4.8 moveto 6 0 rlineto 0 1.5 rlineto -6 0 rlineto 0 -1.5 rlineto fill
0.04 setlinewidth
fgcolor
-3 1.3 moveto 6 0 rlineto stroke
-3 4.8 moveto 6 0 rlineto stroke
-3 6.3 moveto 6 0 rlineto stroke
0.03 setlinewidth
-3 2.3 moveto 6 0 rlineto stroke
-3 3.6 moveto 6 0 rlineto stroke
0.06 setlinewidth
[0.1 0.1] 0.1 setdash
colord
-3 2.3  moveto 6 0 rlineto stroke
colorb
-3 3.6 moveto 6 0 rlineto stroke
[] 0 setdash
initclip
colord
/Width 0.025 def
0 6.3 1.5 90 beglphoton 0.1 90 beglscalar 90 arrow1
1 6.3 1.5 80 beglphoton 0.1 80 beglscalar 80 arrow1
2 6.3 1.5 70 beglphoton 0.1 70 beglscalar 70 arrow1
-1 6.3 1.5 100 beglphoton 0.1 100 beglscalar 100 arrow1
-2 6.3 1.5 110 beglphoton 0.1 110 beglscalar 110 arrow1
fgcolor
[0.1 0.05 0.02 0.05] 0.1 setdash
0.02 setlinewidth
0 0 moveto 3.5 0 lineto stroke
0.4 tb 4 -0.15 moveto (big bang) show
initclip
newpath
0 0 moveto
0 0.5 0.6 1 LL 1.5 curveto 2 5 rlineto
6 0 rlineto currentpoint
pop 0 lineto
0 0 lineto clip
newpath
0 1.3 moveto 3.5 0 rlineto stroke
0.4 tr 4 1.3 0.15 sub moveto (end of inflation) show
newpath
colord
0 2.3 moveto 3.5 0 rlineto stroke
0.4 tb 4 2.3 0.15 sub moveto (EW transition) show
newpath
colorb
0 3.6 moveto 3.5 0 rlineto stroke
0.4 tb 4 3.6 0.15 sub moveto (confinement) show
fgcolor
newpath
0 4.8 moveto 3.5 0 rlineto stroke
0.4 tr 4 4.8 0.15 sub moveto (nucleosynthesis) show
newpath
0 6.3 moveto 3.5 0 rlineto stroke
0.4 tr 4 6.3 0.15 sub moveto (formation of atoms) show
initclip
newpath
[] 0 setdash
1 setlinecap
1 setlinejoin
0.04 setlinewidth
-3.4 -0.1 moveto 0 7.4 rlineto currentpoint stroke 90 arrow1
0.35 tb -4.6 7.5 moveto (time) show
2.6 2.95 moveto -1 0 rlineto currentpoint stroke 180 arrow1
2.75 2.85 moveto 0.35 tbi (Quark Gluon Plasma) show
[0.1 0.05 0.02 0.05] 0.1 setdash
0.02 setlinewidth
initclip
newpath
0 0 moveto
0 0.5 -0.6 1 LL neg 1.5 curveto -2 5 rlineto
-6 0 rlineto currentpoint
pop 0 lineto
0 0 lineto clip
newpath
0 0 moveto -3.3 0 rlineto stroke
newpath
0 1.3 moveto -3.3 0 rlineto stroke
newpath
colord
0 2.3 moveto -3.3 0 rlineto stroke
newpath
colorb
0 3.6 moveto -3.3 0 rlineto stroke
newpath
fgcolor
0 4.8 moveto -3.3 0 rlineto stroke
newpath
0 6.3 moveto -3.3 0 rlineto stroke
initclip
0.35 tr -4.9 1.3 -0.12 add moveto (10) show 0.25 tr 0.02 0.12 rmoveto (-32) show 0.35 tr 0 -0.12 rmoveto (  sec) show
colord
0.35 tb -4.9 2.3 -0.12 add moveto (10) show 0.25 tb 0.02 0.12 rmoveto (-10) show 0.35 tb 0 -0.12 rmoveto (  sec) show
colorb
0.35 tb -4.9 3.6 -0.12 add moveto (10) show 0.25 tb 0.02 0.12 rmoveto (-5) show 0.35 tb 0 -0.12 rmoveto (  sec) show
fgcolor
0.35 tr -4.9 4.8 -0.12 add moveto (10) show 0.25 tr 0.02 0.12 rmoveto (+2) show 0.35 tr 0 -0.12 rmoveto (  sec) show
0.35 tr -4.9 6.3 -0.12 add moveto (10) show 0.25 tr 0.02 0.12 rmoveto (+12) show 0.35 tr 0 -0.12 rmoveto (  sec) show
initclip
newpath
0 0 moveto 
0 0.5 0.6 1 LL 1.5 curveto 3 7.5 rlineto currentpoint
pop 2 mul neg 0 rlineto
3 -7.5 rlineto -0.6 1 0 0.5 0 0 curveto clip
[] 0 setdash
0.015 setlinewidth
1399657368 srand
newpath
150 {
rand int_max div 4 mul -2 add
rand int_max div 2.2 mul 1.35 add
redparton
rand int_max div 4 mul -2 add
rand int_max div 2.2 mul 1.35 add
blueparton
rand int_max div 4 mul -2 add
rand int_max div 2.2 mul 1.35 add
greenparton
} repeat
1399757367 srand
newpath
0.015 setlinewidth
25 {
rand int_max div 4 mul -2 add
rand int_max div 3.7 add
rand int_max div 360 mul
baryon1
rand int_max div 4 mul -2 add
rand int_max div 3.7 add
rand int_max div 360 mul
baryon2
} repeat
1354757367 srand
newpath
0.015 setlinewidth
20 {
rand int_max div 5 mul -2.5 add
rand int_max div 1.3 mul 4.9 add
rand int_max div 360 mul
deuterium1
rand int_max div 5 mul -2.5 add
rand int_max div 1.3 mul 4.9 add
rand int_max div 360 mul
deuterium2
} repeat
1354841379 srand
newpath
0.015 setlinewidth
-4 0.7 4 {
/dd_x exch def
1 {
rand int_max div 1 mul 0.5 sub dd_x add
rand int_max div 1 mul 6.5 add
rand int_max div 360 mul
atom1
} repeat
} for
initclip
1 setlinecap
1 setlinejoin
0.06 setlinewidth
fgcolor
0 0 moveto 
/FF 1.1 def
0 0.5 0.6 1 LL 1.5 curveto 2.12 FF mul 5.3 FF mul rlineto currentpoint stroke
exch neg exch moveto
2.12 FF mul -5.3 FF mul rlineto -0.6 1 0 0.5 0 0 curveto stroke
}
\end{feynman}
\hfil}

\begin{figure}[htbp]
\centerline{
\hfill
\resizebox*{5cm}{!}{\box1}
\hfill
\resizebox*{5cm}{!}{\includegraphics{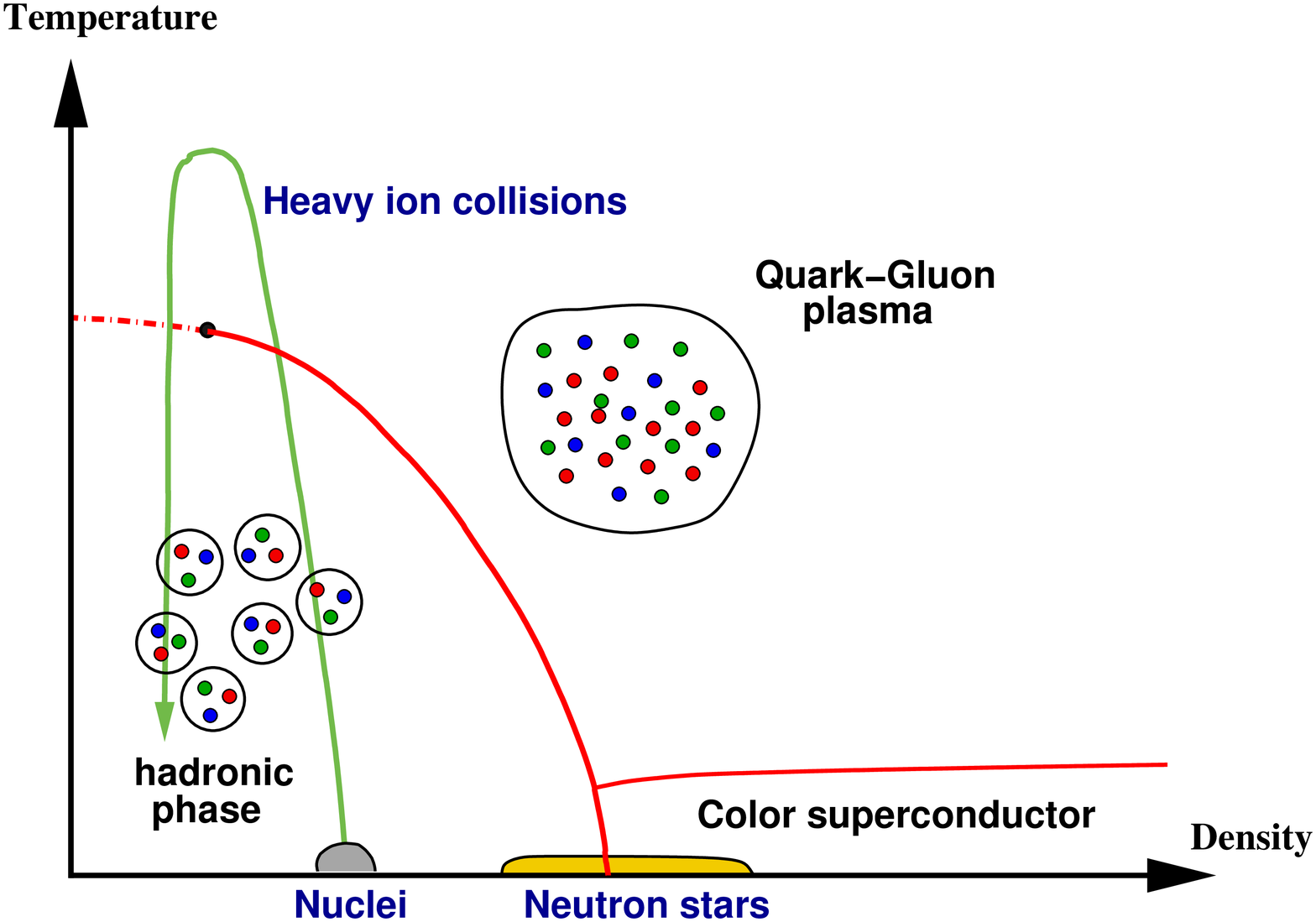}
\hfill
}
}
\caption{\label{fig:phase-diag} Left: quark-gluon plasma in the early universe. Right: QCD phase-diagram.}
\end{figure}
This phase transition has occurred in the expansion of the early
universe (left panel of figure \ref{fig:phase-diag}), but
unfortunately this has not left any visible relic in today's
observable sky. Another place to look for experimental evidence of
color deconfinement is in the collisions of large nuclei at high
energy (see the right panel of figure \ref{fig:phase-diag}). The basic
idea of these experiments is to deposit a large amount of energy in
order to create matter with an energy density larger than the critical
one, and to do so in an extended volume, i.e. large compare to the
typical hadronic size, so that thermodynamical concepts have a chance
to apply. Such collisions have been performed at the AGS (BNL), at the
SPS (CERN), presently at the RHIC (BNL), and in the near future at the
LHC.

\setbox1\hbox to 6cm{
\hfil\begin{feynman}{5cm}
\diagram{25}{
-4.75 1 translate
gsave
newpath
/L 3.25 def
3 0.2 moveto L L rlineto L -2 mul 0 rlineto L neg dup rlineto clip
1 setlinejoin
1 setlinecap
/f 
{/p exch def /M exch def
p 3 add
p p mul M M mul add sqrt 0.2 add
} def
/hyperbole 
{/M exch def
M L neg f moveto
L neg 0.01 L {M exch f lineto} for 
} def
/hyperbole1 
{/M exch def
M L f lineto
L -0.01 L neg {M exch f lineto} for 
} def
0.8441 0.0000 0.0000 setrgbcolor
newpath 0.01 hyperbole 0.3 hyperbole1 fill
1.0000 0.2455 0.0000 setrgbcolor
newpath 0.3 hyperbole 0.6 hyperbole1 fill
1.00 0.75 0.00 setrgbcolor
newpath 0.6 hyperbole 1.7 hyperbole1 fill
0.8664 0.7223 0.5664 setrgbcolor
newpath 1.7 hyperbole 2.5 hyperbole1 fill
0.9664 0.8223 0.666 setrgbcolor
newpath 2.5 hyperbole 5 hyperbole1 fill
0.03 setlinewidth
fgcolor
newpath 0.3 hyperbole stroke
newpath 0.6 hyperbole stroke
newpath 1.65 hyperbole stroke
newpath 1.675 hyperbole stroke
newpath 1.7 hyperbole stroke
newpath 2.5 hyperbole stroke
grestore
fgcolor
0.05 setlinewidth
1 setlinejoin
1 setlinecap
2.3 -0.5 moveto 6.4 3.6 lineto currentpoint stroke 45 arrow
0.05 setlinewidth
3.7 -0.5 moveto -0.4 3.6 lineto currentpoint stroke 135 arrow
0.01 setlinewidth
-1 0.2 moveto 8 0 rlineto currentpoint stroke 0 arrow1
3 -0.4 moveto 0 4.5 rlineto currentpoint stroke 90 arrow1
0.5 ti 7.1 0.1 moveto (z ) show
0.5 ti 3.15 4 moveto (t) show
/max 2147483648 def
/randcoord{rand max div 2.0 mul sqrt rand max div 360 mul
  2 copy cos mul 2 add 3 1 roll sin mul 2 add} def
/nucleon{2 copy 0.5 0 360 arc fill 0 0 0 setrgbcolor 0.5 0 360 arc stroke} def
/proton{0 0 0 setrgbcolor 0.6 setgray nucleon} def
/neutron{0 0 0 setrgbcolor 0.95 setgray nucleon} def
/noyau {
  0.05 setlinewidth
  1298657168 srand
  40 {randcoord proton randcoord neutron} repeat
  2 2 proton
  2.5 3 neutron
} def
newpath
gsave
2.4 -0.95 translate
45 rotate
0.12 0.2 scale
noyau
grestore
gsave
3.27 -0.62 translate
-45 rotate
0.12 0.2 scale
noyau
grestore
fgcolor
0.015 setlinewidth
3.07 0.4 verysmallblob 
moveto 0.3 0 rlineto
3.35 0.35 rlineto currentpoint stroke 
0.33 tr moveto 0.1 -0.07 rmoveto graya (strong fields) show fgcolor
0.1 0.07 rmoveto 0.5 0 rlineto currentpoint 2 copy stroke 0 arrow1
0.33 tr moveto 0.1 -0.07 rmoveto grayd (classical EOMs) show fgcolor
newpath
3.07 0.7 verysmallblob 
moveto 0.6 0 rlineto
1.2 0.7 rlineto currentpoint stroke 
0.33 tr moveto 0.1 -0.07 rmoveto graya (gluons & quarks out of eq.) show fgcolor 
0.1 0.07 rmoveto 0.5 0 rlineto currentpoint 2 copy stroke 0 arrow1
0.33 tr moveto 0.1 -0.07 rmoveto grayd (kinetic theory) show fgcolor
newpath
3.07 1.3 verysmallblob 
moveto 1.2 0 rlineto
1.1 0.75 rlineto currentpoint stroke 
0.33 tr moveto 0.1 -0.07 rmoveto graya (gluons & quarks in eq.) show fgcolor 
0.1 0.0 rmoveto 
0 0.4 rlineto currentpoint 0 0.4 rlineto stroke
moveto 0.5 0 rlineto currentpoint 2 copy stroke 0 arrow1
0.33 tr moveto 0.1 -0.07 rmoveto grayd (hydrodynamics) show fgcolor
newpath
3.07 2.3 verysmallblob 
moveto 2.2 0 rlineto
1.25 0.4 rlineto currentpoint stroke 
0.33 tr moveto 0.1 -0.07 rmoveto graya (hadrons in eq.) show fgcolor
newpath
3.07 3.1 verysmallblob 
moveto 3 0 rlineto
0.8 0.2 rlineto currentpoint stroke 
0.33 tr moveto 0.1 -0.07 rmoveto colora (freeze out) show  fgcolor
newpath    
fgcolor
3 0.2 blob pop pop
}
\end{feynman}\hfil}
\begin{figure}[htbp]
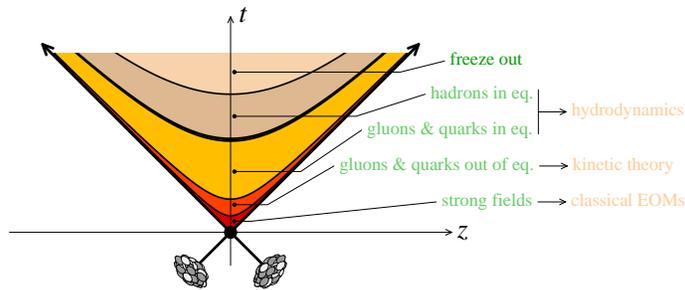

\centerline{
\resizebox*{5cm}{!}{\box1}
}
\caption{\label{fig:stages}Successive stages of the collision of two nuclei.}
\end{figure}
The standard scenario for a high energy nucleus-nucleus collision
involves several stages (see figure \ref{fig:stages}). At extremely
short time scales take place the very hard processes that account for
the hard particles in the final state. It is believed that they can be
calculated using the standard tools of perturbative QCD and collinear
factorization. However, most of the particles that make the final
state are in fact quite soft (99\% of the multiplicity in a collision
at RHIC is made of particles with $p_\perp\le 2$~GeV). The bulk of
this particle production takes place slightly later ($t\sim 0.2$~fm),
and since it involves the low $x$ part of the nuclear wavefunction,
i.e. large parton densities, it is expected to be amenable to a
treatment in terms of classical fields. The Color Glass Condensate
\cite{cgc} effective theory is a framework in which such calculations
can be carried out in a systematic way. Eventually, this deconfined
matter reaches a state of local thermal equilibrium, and can be
described by hydrodynamics. Because it is in expansion, it cools down
and reaches the critical temperature where hadrons are formed
again. At a later stage, the density becomes too low to have an
interaction rate high enough to sustain equilibrium, and the system
freezes out. Experimentally, one is trying to infer properties of the
early and intermediate stages of these collisions from the measured
hadrons in the final state.

\section{Small coupling wisdom}
Until recently, the theoretical studies of the properties of the quark
gluon plasma have been based on the assumption that the temperature is
high enough to apply weak coupling techniques, thanks to asymptotic
freedom (the average distance between two particles in the plasma goes
like $T^{-1}$). When the coupling constant $g$ is small, there is a
useful separation between various distance scales, each of them
corresponding to different physical phenomena, and being described by
a specific effective theory \cite{htl}. These scales are~:
\begin{itemize}
\item $\ell\sim T^{-1}$. This is the mean distance between two
particles in the plasma. Also, $T$ is the typical energy of a plasma
particle, and it is therefore this scale that dominates the bulk of
the pressure ($P\sim T^4$).
\item $\ell\sim (gT)^{-1}$. This is the scale at which the main
collective effects occur in the QGP. For instance, the Debye screening
length is of the order of $(gT)^{-1}$, and plasma particles acquire a
thermal mass of order $gT$ due to their interactions with the other
particles.
\item $\ell\sim (g^2T\ln(1/g))^{-1}$ is the mean free path between two
collisions with a soft momentum transfer, i.e. a scattering angle of
order $g$. This is the relevant scale for phenomena such as
electromagnetic emission from the plasma.
\item $\ell\sim (g^4T)^{-1}$ is the mean free path between two hard
scatterings, i.e. with a scattering angle of order unity. This is the
natural scale for the transport of momentum.
\end{itemize}
However, one also sees from this small coupling analysis that the
scale that characterizes thermalization is the last one,
i.e. $(g^4T)^{-1}$. In other words, it is quite unnatural to have a
fast thermalization in a weakly coupled system.

\section{RHIC results}
In seven years of operation, the RHIC has obtained many results
relative to heavy ion collisions \cite{exp}. Let me only discuss two
specific results, that are amongst the most important.  \setbox2\hbox
to 6cm{\hfil
\begin{feynman}{4.5cm}
\diagram{16}{
  -2 2.5 translate
  1 setlinejoin
  [] 0 setdash
  1 setlinecap
  0 setgray
  /arw {
    /ll exch def
    /yy exch def
    /xx exch def
    0.9 setgray
    xx yy moveto
    0 0.2 rlineto
    ll 0 rlineto
    0 0.2 rlineto
    ll 0.3 mul -0.4 rlineto
    ll 0.3 mul neg -0.4 rlineto
    0 0.2 rlineto
    ll neg 0 rlineto
    0 0.2 rlineto fill
    0 setgray
    0.07 setlinewidth
    xx yy moveto
    0 0.2 rlineto
    ll 0 rlineto
    0 0.2 rlineto
    ll 0.3 mul -0.4 rlineto
    ll 0.3 mul neg -0.4 rlineto
    0 0.2 rlineto
    ll neg 0 rlineto
    0 0.2 rlineto
    stroke
  } def
  initclip
  newpath -1 0.35 moveto 6.9 0 rlineto 0 2.4 rlineto -6.9 0 rlineto closepath
  gsave clip newpath
  << /ShadingType 2
    /ColorSpace /DeviceRGB
    /Coords [0 0.35 0 2.75]
    /AntiAlias true 
    /Function << /FunctionType 2
                 /Domain [0 1]
                 /C0 [0 0 0]
                 /C1 [0.7 0.8 0.8]
                 /N 0.6
              >>
  >> shfill 
  initclip
  newpath -1 0.35 moveto 6.9 0 rlineto 0 -2.4 rlineto -6.9 0 rlineto closepath
  gsave clip newpath
  << /ShadingType 2
    /ColorSpace /DeviceRGB
    /Coords [0 0.35 0 -2.05]
    /AntiAlias true 
    /Function << /FunctionType 2
                 /Domain [0 1]
                 /C0 [0 0 0]
                 /C1 [0.7 0.8 0.8]
                 /N 0.6
              >>
  >> shfill 
  initclip
  gsave
  -1 0.35 translate
  0 0 -1 arw
  0.2 1 scale
  1 1 1 setrgbcolor
  0 0 2.4 0 360 arc fill
  initclip newpath 0 0 2.4 0 360 arc clip
  fgcolor
  newpath 170 {0 0 moveto
              rand int_max div dup mul 1 exch sub dup dup mul /RR exch def 2.1 mul
              rand int_max div 360 mul polto
	      1 RR sub rand int_max div dup mul mul RR add
	      1 RR sub rand int_max div dup mul mul RR add
	      1 RR sub rand int_max div dup mul mul RR add
	      setrgbcolor 
              newpath 0.1 5 blob2 pop pop} repeat
  0 setgray
  0.08 setlinewidth
  0 0 2.4 0 360 arc stroke
  grestore
  gsave
  5.9 0.35 translate
  0 0 +1 arw
  0.2 1 scale
  1 1 1 setrgbcolor
  0 0 2.4 0 360 arc fill
  initclip newpath 0 0 2.4 0 360 arc clip
  fgcolor
  newpath 170 {0 0 moveto
              rand int_max div dup mul 1 exch sub dup dup mul /RR exch def 2.1 mul
              rand int_max div 360 mul polto
	      1 RR sub rand int_max div dup mul mul RR add
	      1 RR sub rand int_max div dup mul mul RR add
	      1 RR sub rand int_max div dup mul mul RR add
	      setrgbcolor 
              newpath 0.1 5 blob2 pop pop} repeat
  0 setgray
  0.08 setlinewidth
  0 0 2.4 0 360 arc stroke
  grestore
  1.0000 0.2455 0.0000 setrgbcolor
  /Width 0.05 def
  newpath
  -1 1.8 blob
  4 0 beglphoton
  2 copy
  1 20 beglphoton 2 copy
  5.9 2.1 getlparam beglphoton blob pop pop
  <<
   /PatternType 2
   /Shading <<
   /ShadingType 2
   /ColorSpace /DeviceRGB
   /Coords [3 -1.6 6 6.7]
   /Function <<
     /FunctionType 2
     /Domain [0 1]
     /C0 [1 1 1]
     /C1 [1.0000 0.2455 0.0000]
     /N 0.7
     >>
   >>
  >>
  matrix
  makepattern setpattern
  %
  /branch {
    counttomark 0 gt
    {
    dup lmax lt {newbranch} {pop arrow1} ifelse
    branch
    }
    {} ifelse
  } def
  /newbranch {
    /tmp_l exch def
    /tmp_th exch def
    /tmp_y exch def
    /tmp_x exch def
    /tmp_dl tmp_l Length def
    /tmp_dth tmp_l Opening 2 div def
    rand int_max div 0.45 gt
    {
    newpath
    tmp_x tmp_y tmp_dl tmp_th tmp_dth add
    tmp_dl 0.8 lt {beglscalar}{beglphoton} ifelse 
    tmp_th tmp_dth add tmp_l 1 add
    newpath
    tmp_x tmp_y tmp_dl tmp_th tmp_dth sub
    tmp_dl 0.8 lt {beglscalar}{beglphoton} ifelse
    tmp_th tmp_dth sub tmp_l 1 add
    }
    {
    tmp_x tmp_y tmp_th tmp_l 1 add
    }
    ifelse
  } def
  /Length {
    /tmp_l exch def
    0.85 tmp_l 0.5 add 0.6 exp div
  } def
  /Opening {
    /tmp_l exch def
    10 rand int_max div 10 mul add
  } def	
  /Width 0.04 def
  1 60 beglphoton
  /lmax 8 def
  1234869 srand
  60 0 mark 5 1 roll branch
  pop
  /Length {
    /tmp_l exch def
    0.55 tmp_l 0.5 add 0.6 exp div
  } def
  /Opening {
    /tmp_l exch def
    30 rand int_max div 40 mul add
  } def	
  /Width 0.04 def
  0.7 -110 beglphoton
  /lmax 10 def
  1237869 srand
  -110 0 mark 5 1 roll branch
}
\end{feynman}
\hfil
}

\begin{figure}[htbp]
\centerline{
\hfill
\resizebox*{5cm}{!}{\includegraphics{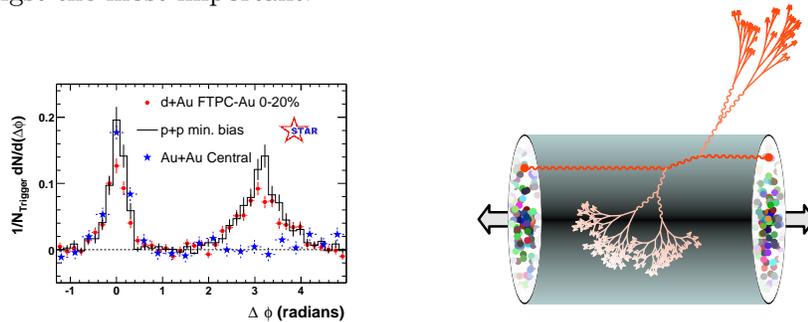}}
\hfill
\resizebox*{5cm}{!}{\box2}
\hfill
}
\caption{\label{fig:quenching}Left: azimuthal correlation at
RHIC. Right: cartoon of jet quenching.}
\end{figure}
The first one is the phenomenon of {\sl jet quenching}. In
proton-proton collisions, one is used to final states being
predominantly 2-jet events. Jets are very hard to measure directly in
nuclear collisions due to the high final state multiplicity, but one
can measure the azimuthal correlation function between pairs of
particles (see the left panel of figure \ref{fig:quenching}). In pp
collisions, this correlation function is peaked both at $\phi=0$ and
at $\phi=\pi$. The same is true in deuteron-nucleus
collisions. However, in nucleus-nucleus collisions, one observes only
a correlation around $\phi=0$, and no correlation at $\phi=\pi$. This
led to the idea that the matter formed in nuclear collisions is opaque
to the propagation of hard partons (i.e. they fragment much more than
in the vacuum and are undistinguishable from the bulk of the other
particles when they come out on the other side). The correlation at
$\phi=0$ would be from jets produced near the surface of the medium, a
configuration in which the partner jet in the opposite direction has a
much larger length of matter to cross before escaping, hence its
disappearance.

\setbox1\hbox to 7cm{\hfil
\begin{feynman}{3.3cm}
\diagram{16}{
  -3 3 translate
  1 setlinejoin
  [] 0 setdash
  1 setlinecap
  0 setgray
  1111 srand
  /arw {
    /ll exch def
    /yy exch def
    /xx exch def
    0.9 setgray
    xx yy moveto
    0 0.2 rlineto
    ll 0 rlineto
    0 0.2 rlineto
    ll 0.3 mul -0.4 rlineto
    ll 0.3 mul neg -0.4 rlineto
    0 0.2 rlineto
    ll neg 0 rlineto
    0 0.2 rlineto fill
    0 setgray
    0.07 setlinewidth
    xx yy moveto
    0 0.2 rlineto
    ll 0 rlineto
    0 0.2 rlineto
    ll 0.3 mul -0.4 rlineto
    ll 0.3 mul neg -0.4 rlineto
    0 0.2 rlineto
    ll neg 0 rlineto
    0 0.2 rlineto
    stroke
  } def
  initclip
  newpath -1 0.75 moveto 4.9 0 rlineto 0 1.65 rlineto -4.9 0 rlineto closepath
  gsave clip newpath
  << /ShadingType 2
    /ColorSpace /DeviceRGB
    /Coords [0 0.75 0 2.4]
    /AntiAlias true 
    /Function << /FunctionType 2
                 /Domain [0 1]
                 /C0 [0 0 0]
                 /C1 [0.7 0.8 0.8]
                 /N 0.6
              >>
  >> shfill 
  initclip
  newpath -1 0.75 moveto 4.9 0 rlineto 0 -1.65 rlineto -4.9 0 rlineto closepath
  gsave clip newpath
  << /ShadingType 2
    /ColorSpace /DeviceRGB
    /Coords [0 0.75 0 -1.1]
    /AntiAlias true 
    /Function << /FunctionType 2
                 /Domain [0 1]
                 /C0 [0 0 0]
                 /C1 [0.7 0.8 0.8]
                 /N 0.6
              >>
  >> shfill 
  initclip
  gsave
  -1 1.5 translate
  0 0 -1 arw
  0.2 1 scale
  1 1 1 setrgbcolor
  0 0 2.4 0 360 arc fill
  initclip newpath 0 0 2.4 0 360 arc clip
  fgcolor
  newpath 170 {0 0 moveto
              rand int_max div dup mul 1 exch sub dup dup mul /RR exch def 2.1 mul
              rand int_max div 360 mul polto
	      1 RR sub rand int_max div dup mul mul RR add
	      1 RR sub rand int_max div dup mul mul RR add
	      1 RR sub rand int_max div dup mul mul RR add
	      setrgbcolor 
              newpath 0.1 5 blob2 pop pop} repeat
  0 setgray
  0.08 setlinewidth
  0 0 2.4 0 360 arc stroke
  grestore
  gsave
  3.9 0 translate
  0 0 +1 arw
  0.2 1 scale
  1 1 1 setrgbcolor
  0 0 2.4 0 360 arc fill
  initclip newpath 0 0 2.4 0 360 arc clip
  fgcolor
  newpath 170 {0 0 moveto
              rand int_max div dup mul 1 exch sub dup dup mul /RR exch def 2.1 mul
              rand int_max div 360 mul polto
	      1 RR sub rand int_max div dup mul mul RR add
	      1 RR sub rand int_max div dup mul mul RR add
	      1 RR sub rand int_max div dup mul mul RR add
	      setrgbcolor 
              newpath 0.1 5 blob2 pop pop} repeat
  0 setgray
  0.08 setlinewidth
  0 0 2.4 0 360 arc stroke
  grestore
  gsave
  8.5 0 translate
  initclip newpath 0 0 2.4 0 360 arc clip
  newpath 500 {0 0 moveto
    rand int_max div dup mul 1 exch sub dup dup mul /RR exch def 2.1 mul
    rand int_max div 360 mul polto
    1 RR sub rand int_max div dup mul mul RR add
    1 RR sub rand int_max div dup mul mul RR add
    1 RR sub rand int_max div dup mul mul RR add
    setrgbcolor 
    newpath 0.1 0 360 arc fill} repeat
  initclip newpath 0 1.5 2.4 0 360 arc clip
  newpath 500 {0 1.5 moveto
    rand int_max div dup mul 1 exch sub dup dup mul /RR exch def 2.1 mul
    rand int_max div 360 mul polto
    1 RR sub rand int_max div dup mul mul RR add
    1 RR sub rand int_max div dup mul mul RR add
    1 RR sub rand int_max div dup mul mul RR add
    setrgbcolor 
    newpath 0.1 0 360 arc fill} repeat
  gsave
  0 0.75 translate
  1 0.75 scale
  initclip
  newpath 
  0 0 2.15 0 360 arc gsave clip newpath
  << /ShadingType 3
    /ColorSpace /DeviceRGB
    /Coords [0 0 0 0 0 2.15]
    /AntiAlias true 
    /Function << /FunctionType 2
                 /Domain [0 1]
                 /C0 [0 0 0]
                 /C1 [0.7 0.8 0.8]
                 /N 0.6
              >>
  >> shfill 
  grestore
  grestore
  initclip
  gsave
  -1 0.4 translate
  fgcolor
  /R 0.35 def
  /N 7 def
  0.05 setlinewidth
  1 1 N {
    /i exch def
    0 0 moveto R 360 N div i mul polto lineto currentpoint stroke
    360 N div i mul arrow1
  } for
  [0.07 0.07] 0 setdash
  0 0 R 0.05 add 0 360 arc stroke
  [] 0 setdash
  grestore
  gsave
  -1 0.4 translate
  0.6 1.4 scale
  fgcolor
  /R0 0.35 def
  /R 0.8 def
  /N 7 def
  0.05 setlinewidth
  1 1 N {
    /i exch def
    0 0 moveto R 360 N div i mul polto moveto
    R0 360 N div i mul polto lineto
    currentpoint stroke
    360 N div i mul arrow1
  } for
  [0.07 0.07] 0 setdash
  0 0 R R0 add 0.05 add 0 360 arc stroke
  [] 0 setdash
  grestore
  gsave
  -1 0.4 translate
  fgcolor
  0.02 setlinewidth
  [0.02 0.1] 0 setdash
  0 0 R R0 add 0.05 add 0 360 arc stroke
  [] 0 setdash
  grestore
  fgcolor
  0.05 setlinewidth
  0 0   2.4 0 360 arc stroke
  0 1.5 2.4 0 360 arc stroke
  grestore
}
\end{feynman}
\hfil}

\begin{figure}[htbp]
\centerline{\hskip -5mm
\resizebox*{4.5cm}{!}{\includegraphics{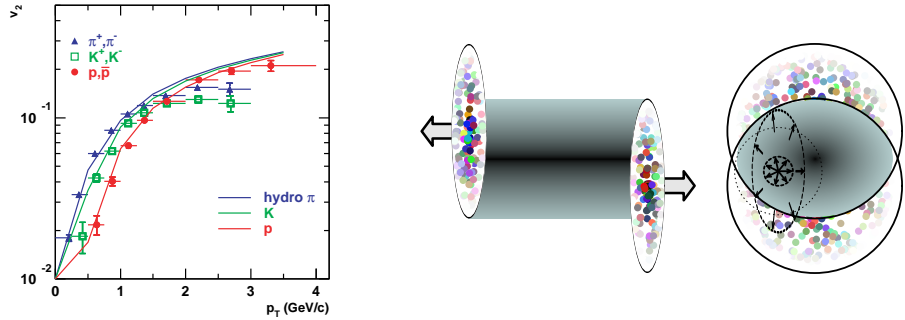}}
\hfil
\resizebox*{6cm}{!}{\box1}
\hfil
}
\vskip 2mm
\caption{\label{fig:v2}Left: elliptic flow $v_2$ measured at RHIC,
compared with hydrodynamical simulations. Right: cartoon of a
peripheral AA collision.}
\end{figure}
The second striking phenomenon observed at RHIC is {\sl elliptic
flow}. In peripheral collisions, the initial shape of the matter
formed in the collision has an elliptic azimuthal shape (see figure
\ref{fig:v2}). Since its pressure is zero at the outer surface, and
maximal at the center, the pressure gradients are larger in the
direction of the small axis of the ellipsis. This means that the
particles will acquire a larger flow velocity in this direction, and
the net result of this is an anisotropy of the transverse momenta of
the particles measured in the final state. This anisotropy is
characterized by the measure of a number called $v_2$, defined by
\begin{equation}
\frac{dN}{d\phi}\sim v_2\cos(2(\phi-\Phi_{_R}))\; ,
\end{equation}
where $\Phi_{_R}$ is the azimuthal direction of the reaction plane,
i.e. the plane that contains the impact parameter of the
collision. Experimental results for $v_2$ are displayed for various
particles, as a function of their $p_\perp$, in the left panel of
figure \ref{fig:v2}. The solid curves in this plot are the result of a
simulation based on ideal hydrodynamics, in which one treats the
matter produced in the collision as a fluid that has no viscosity. Let
us recall here that no viscosity means that the fluid must be in
equilibrium so that it is not the siege of dissipative phenomena.
Moreover, it has been argued that the system must have been close to
equilibrium from early times, in order to produce an elliptic flow of
that magnitude (dissipative effects tend to reduce $v_2$), which is
hard to explain in the conventional weak coupling approach.

\section{AdS/CFT duality and the QGP}
The parameter that controls viscous effects in hydrodynamics is the
dimensionless ratio of the viscosity to entropy density,
$\eta/s$. When evaluated in perturbative QCD, this ratio is
$\eta/s\sim (g^4\ln(1/g))^{-1}$ \cite{eta}. It is therefore large in
the weak coupling limit, in apparent contradiction with the fact that
non viscous hydrodynamics reproduces well RHIC data on $v_2$. $\eta/s$
cannot be evaluated in the strong coupling limit of QCD. However, one
can perform such strong coupling calculations in an ${\cal N}=4$
super-symmetric Yang-Mills theory, thanks to the AdS/CFT
correspondence.  \setbox1\hbox to 5cm{ \hfil
  \begin{feynman}{3.5cm}
    \diagram{15}{
      initstate
      -6 1.5 translate
      fgcolor
      0.07 setlinewidth
      4 5 moveto
      0 -6 rlineto
      currentpoint
      stroke
      -90 arrow
      gsave
      0 0 translate
      initclip
      newpath 
      0 0 moveto
      6 0 rlineto
      3 2 rlineto
      -6 0 rlineto
      closepath gsave clip newpath
      << /ShadingType 2
         /ColorSpace /DeviceRGB
         /Coords [0 0 0 2]
         /AntiAlias true 
         /Function << /FunctionType 2
                 /Domain [0 1]
                 /C0 [0 0 0]
                 /C1 [0.8 0.8 0.8]
                 /N 0.6
              >>
      >> shfill
      grestore initclip
      0.06 setlinewidth
      fgcolor
      0 0 moveto
      6 0 rlineto
      3 2 rlineto
      -6 0 rlineto
      -3 -2 rlineto stroke
      gsave
      2.3 2.6 blob
      /Width 0.06 def
      1.4 -200 10 begcphoton
      blob
      gsave
      0 1.8 translate
      initclip
      newpath 
      0 0 moveto
      6 0 rlineto
      3 2 rlineto
      -6 0 rlineto
      closepath gsave clip newpath
      << /ShadingType 2
         /ColorSpace /DeviceRGB
         /Coords [0 0 0 2]
         /AntiAlias true 
         /Function << /FunctionType 2
                 /Domain [0 1]
                 /C0 [0 0 0]
                 /C1 [0.9 0.8 0.7]
                 /N 0.6
              >>
      >> shfill
      grestore initclip
      0.06 setlinewidth
      fgcolor
      0 0 moveto
      6 0 rlineto
      3 2 rlineto
      -6 0 rlineto
      -3 -2 rlineto stroke
      grestore
      fgcolor
      2.3 2.6 blob
      /Width 0.005 def
      1.4 -200 10 begcphoton
      blob
      fgcolor
      0.07 setlinewidth
      4 5 moveto
      4 2.8 lineto stroke
      4 1.8 moveto
      4 1 lineto stroke
      0.015 setlinewidth
      3.5 2.8 moveto 6 0 rlineto currentpoint stroke 0 arrow1
      0.015 setlinewidth
      3.4 2.4 moveto 3 2 rlineto currentpoint stroke 2 3 atan arrow1
      0.6 tbi
      4.3 -1 moveto (z) show
      9.9 2.6 moveto (x) show
      -0.35 0.4 rmoveto 0.5 0 rlineto currentpoint stroke 0 arrow1
      6.6 4.5 moveto (t) show
      0.5 tr 9.3 3.7 moveto (our world : ) show 0.5 tbi (z = 0) show
      0.5 tr 9.3 1.8 moveto (horizon : ) show 0.5 tbi (z = ) show
      0.5 tb (1/) show
      0.5 symb (p) show 0.5 tbi (T) show
    }
  \end{feynman}
  \hfil
}

\setbox2\hbox to 7cm{
  \hfil
  \begin{feynman}{3cm}
    \diagram{15}{
      initstate
      -3 1 translate
      fgcolor
      0.05 setlinewidth
      0 0 moveto 10 0 lineto currentpoint stroke 0 arrow
      0 0 moveto 0 7 lineto currentpoint stroke 90 arrow
      0.6 tbi
      10.5 -0.2 moveto (gN) show
      0.4 tbi 0.07 0.2 rmoveto (2) show
      0.6 symb
      -0.4 7.3 moveto (h / ) show 0.6 tbi (s) show
      0.01 setlinewidth
      [0.01 0.1] 0 setdash
      0 1 moveto 10 0 rlineto stroke
      0.5 tb -1.5 0.85 moveto (1 / 4) show 0.5 symb (p) show
      newpath initclip
      0 0 moveto 10 0 rlineto 0 6.6 rlineto -10 0 rlineto 0 -6.6 rlineto clip
      newpath
      [0.03 0.1] 0 setdash
      /f {
	/x exch def
	1 x div 1 add 
      } def
      newpath
      0.1 dup f moveto
      0.1 0.03 10 {
	dup f lineto
      } for
      stroke
      [] 0 setdash
      0.1 setlinewidth
      newpath
      colorb
      0.1 dup f moveto
      0.1 0.03 1 {
	dup f lineto
      } for
      stroke
      newpath
      colord
      5 dup f moveto
      5 0.03 10 {
	dup f lineto
      } for
      stroke
      fgcolor
      0.02 setlinewidth
      7.5 2.3 moveto 0 -1 rlineto currentpoint stroke -90 arrow1
      0.55 tb 5.8 2.5 moveto (AdS/CFT duality) show
      1.4 5.5 moveto -1 0 rlineto currentpoint stroke 180 arrow1
      0.55 tb 1.6 5.4 moveto (perturbation theory) show
    }
  \end{feynman}
  \hfil
}
\begin{figure}[htbp]
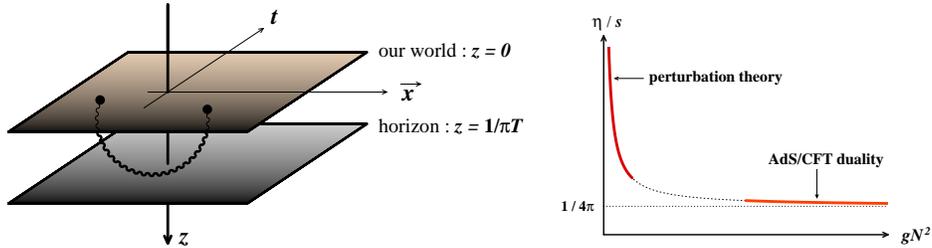

\centerline{
\hfill
\resizebox*{5cm}{!}{\box1}
\hfill
\resizebox*{5cm}{!}{\box2}
\hfill
}
\vskip 2mm
\caption{\label{fig:adscft}Left: AdS/CFT setup for a thermal system. Right: $g$ dependence of $\eta/s$.}
\end{figure}
The AdS/CFT correspondence states that this particular gauge theory is
dual to a type IIB string theory on an $AdS_5\times S_5$
background. Moreover, the correspondence between the parameters of the
two theories is such that the limit $g^2\ll 1, g^2 N_c\gg 1$ of the
gauge theory corresponds to the limit where the string theory
simplifies into classical super-gravity. Then, correlators in the gauge
theory can be calculated from classical solutions in the gravity dual,
with boundary conditions on the 4-dimensional boundary of the $AdS_5$
that depend on the specific problem one wants to study in the
4-dimensional gauge theory.

Using this correspondence, $\eta/s$ was calculated exactly in the
strong coupling limit of an ${\cal N}=4$ super-symmetric Yang-Mills
theory, and it was found to be $\eta/s=1/4\pi$ \cite{ads}. It was also
argued that $\eta/s$ must have a lower bound of order unity because of
the uncertainty principle ($\eta/s$ is of the order of the ratio of
the mean free path to the De Broglie wavelength of the particles), and
that the ${\cal N}=4$ SUSY Yang-Mills theory in the strong coupling
limit actually realizes the lower bound. Although one cannot use these
techniques in QCD, it is believed that the strong coupling regime of
QCD also exhibits a small value of $\eta/s$, but it is at the moment
impossible to make this statement more rigorous than an educated
guess.

\section{Early time dynamics}
As we have seen, invoking a strong gauge coupling may explain why the
viscosity is so small, and also why the system thermalizes
early. Another point of view exists in the community, which assumes
that the early stages of high energy heavy ion collisions can be
studied by perturbative techniques. First of all, one should recall
that the bulk of particle production in heavy ion collisions is due to
partons that carry a small momentum fraction $x$ in the incoming
nuclei. Because of the rise of the gluon distribution at small $x$,
the density of such partons is high, leading to the phenomenon of
saturation: the gluon phase-space density cannot grow larger than
$1/\alpha_s$ due to the repulsive interactions among the gluons. Thus,
the gluons at low $x$ must occupy higher $p_\perp$ modes: at a given
$x$, all the modes up to $p_\perp=Q_{\rm sat}(x)$ are occupied with a
phase-space density of the order of $1/\alpha_s$. $Q_{\rm sat}(x)$,
the saturation momentum, grows like $x^{-0.3}$ at small $x$. It also
depends on the nuclear atomic number like $A^{1/3}$, which means than
nuclei have a larger saturation momentum than protons at a given $x$.

\setbox1\hbox to 8cm{
\hfil\begin{feynman}{6.5cm}
\diagram{25}{
-4.5 0.7 translate
1 setlinejoin
1 setlinecap
0.847 0.72 0.525 setrgbcolor
0 0 moveto 5 6 lineto 0 6 lineto 0 0 lineto fill
colorb
0.06 setlinewidth
0 0 moveto 5 6 lineto stroke
colorf
0 0 moveto 0.5 0 rlineto 0 6 rlineto -0.5 0 rlineto 0 -6 rlineto fill
0.03 setlinewidth
fgcolor
0.5 0 moveto 0 6 rlineto stroke
initclip 2 1 0.5 0 360 arc clip
1398657368 srand
newpath
0.01 setlinewidth
1.75 0.85 bigredparton
2.25 0.9 bigblueparton
2.1 1.25 biggreenparton
initclip
fgcolor
0.03 setlinewidth
2 1 0.5 0 360 arc stroke
initclip 5 1 0.5 0 360 arc clip
1398657168 srand
newpath
0.01 setlinewidth
20 {
rand int_max div 4.5 add
rand int_max div 0.5 add
smallredparton
rand int_max div 4.5 add
rand int_max div 0.5 add
smallblueparton
rand int_max div 4.5 add
rand int_max div 0.5 add
smallgreenparton
} repeat
initclip
fgcolor
0.03 setlinewidth
5 1 0.5 0 360 arc stroke
initclip 2 5 0.5 0 360 arc clip
1398657168 srand
newpath
0.01 setlinewidth
50 {
rand int_max div 1.5 add
rand int_max div 4.5 add
bigredparton
rand int_max div 1.5 add
rand int_max div 4.5 add
bigblueparton
rand int_max div 1.5 add
rand int_max div 4.5 add
biggreenparton
} repeat
initclip
fgcolor
0.03 setlinewidth
2 5 0.5 0 360 arc stroke
initclip
fgcolor
0.09 setlinewidth
0 0 moveto 6 0 rlineto currentpoint stroke 0 arrow
0 0 moveto 0 6.3 rlineto currentpoint stroke 90 arrow 
0.4 tr
6.4 -0.1 moveto (log\noexpand\() show
0.4 ti
(Q) show
0.27 tr
0.08 0.15 rmoveto (2) show
0.4 tr
0 -0.15 rmoveto
(\noexpand\)) show
0.4 tr
-1.6 5.8 moveto (log\noexpand\() show
0.4 ti (x) show
0.27 tr
0.05 0.15 rmoveto (-1) show
0 -0.15 rmoveto
0.4 tr (\noexpand\)) show
0.4 symb
0.3 -0.5 moveto (L) show
0.25 tr
0 -0.15 rmoveto (QCD) show
}
\end{feynman}\hfil
}

\setbox2\hbox to 5cm{\hfil
\begin{feynman}{3.0cm}
\diagram{16}{
  -3 3.00 translate
  1 setlinejoin
  [] 0 setdash
  1 setlinecap
  0 setgray
  /arw {
    /ll exch def
    /yy exch def
    /xx exch def
    0.9 setgray
    xx yy moveto
    0 0.2 rlineto
    ll 0 rlineto
    0 0.2 rlineto
    ll 0.3 mul -0.4 rlineto
    ll 0.3 mul neg -0.4 rlineto
    0 0.2 rlineto
    ll neg 0 rlineto
    0 0.2 rlineto fill
    0 setgray
    0.07 setlinewidth
    xx yy moveto
    0 0.2 rlineto
    ll 0 rlineto
    0 0.2 rlineto
    ll 0.3 mul -0.4 rlineto
    ll 0.3 mul neg -0.4 rlineto
    0 0.2 rlineto
    ll neg 0 rlineto
    0 0.2 rlineto
    stroke
  } def
  -1 0.35 -1 arw
  gsave
  -1 0.35 translate
  0.2 1 scale
  0.2713 0.9684 0.2986 setrgbcolor
  0 0 2.4 0 360 arc fill
  0 setgray
  0.08 setlinewidth
  0 0 2.4 0 360 arc stroke
  grestore
  6.9 0.35 +1 arw
  gsave
  6.9 0.35 translate
  0.2 1 scale
  1.0000 0.7901 0.5563 setrgbcolor
  0 0 2.4 0 360 arc fill
  0 setgray
  0.08 setlinewidth
  0 0 2.4 0 360 arc stroke
  grestore
  6.9 2 blob 2 190 beglphoton 2 copy 
  6.8 1 getlparam beglphoton blob pop pop
  1 170 beglphoton 2 copy
  1 170 beglphoton 2 copy
  1.5 60 beglphoton 0.2 60 beglscalar 60 arrow1
  2 180 beglphoton 2 copy
  -1 1.3 getlparam beglphoton blob pop pop
  -0.9 2.2 getlparam beglphoton blob pop pop
  1 200 beglphoton 2 copy
  1.4 -60 beglphoton 0.2 -60 beglscalar -60 arrow1
  1 200 beglphoton 2 copy
  1.3 -120 beglphoton 0.2 -120 beglscalar -120 arrow1
  -1.1 0.8 getlparam beglphoton blob pop pop
  colord
  6.8 0 blob
  2 160 beglphoton 2 copy 2 copy
  6.9 0.7 getlparam beglphoton blob pop pop
  2 100 beglphoton 0.2 100 beglscalar 100 arrow1
  1 180 beglphoton 2 copy
  1.8 -90 beglphoton 0.2 -90 beglscalar -90 arrow1
  1 170 beglphoton 2 copy
  1 140 beglphoton 2 copy
  1.6 120 beglphoton 0.2 120 beglscalar 120 arrow1
  1 180 beglphoton 2 copy
  -1 1.8 getlparam beglphoton blob pop pop
  1 210 beglphoton 2 copy
  1.6 -110 beglphoton 0.2 -110 beglscalar -110 arrow1
  -1 0.4 getlparam beglphoton blob pop pop
  1.5 -135 beglphoton 0.2 -135 beglscalar -135 arrow1
  grayc
  6.9 -1.5 blob
  2 170 beglphoton 2 copy
  6.9 -0.9 getlparam beglphoton blob pop pop
  1.5 160 beglphoton 2 copy
  7 -0.4 getlparam beglphoton blob pop pop
  1 180 beglphoton 2 copy
  1.5 200 beglphoton 2 copy
  -1 -1.5 getlparam beglphoton blob pop pop
  -1.1 -0.9 getlparam beglphoton blob pop pop
  2 170 beglphoton 2 copy
  -0.9 0 getlparam beglphoton blob pop pop
  -1 -0.4 getlparam beglphoton blob pop pop
}
\end{feynman}
\hfil}
\begin{figure}[htbp]
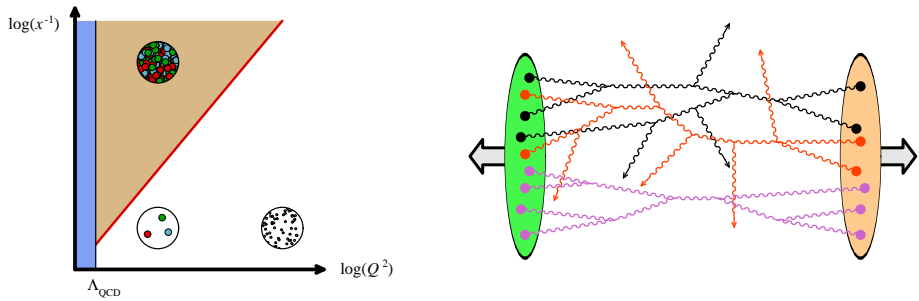

\centerline{
\hfill
\resizebox*{5cm}{!}{\box1}
\hfill
\resizebox*{5cm}{!}{\box2}
\hfill
}
\vskip 2mm
\caption{\label{fig:sat}Left: saturation domain. Right: typical particle production process in the Color Glass Condensate approach.}
\end{figure}
In the saturated regime ($Q^2\le Q_{\rm sat}^2(x)$ in the diagram on
the left of figure \ref{fig:sat}), gluon recombination and multiple
scatterings play a crucial role in the mechanisms of particle
production (see the right panel of figure \ref{fig:sat}). The Color
Glass Condensate \cite{cgc} is an effective theory in which these
effects are systematically taken into account, based on the separation
of the degrees of freedom into color fields (that represent the low
$x$ partons), and static color sources (that represent the large $x$
partons). In particular, the single inclusive spectrum of the gluons
and quarks produced in the collision of two heavy ions have been
computed at leading order \cite{cgc1} in this framework. Work is under
way in order to evaluate the NLO corrections to the gluon spectrum,
and to establish a factorization theorem for nucleus-nucleus
collisions in the saturated regime \cite{cgc1}.  Note that these
calculations only give the distribution of produced particles at a
short time after the impact (of the order of $\tau\sim Q_{\rm
sat}^{-1}$), which can then be used in order to model the initial
conditions for the subsequent hydrodynamical evolution.

\setbox1\hbox to
5cm{\hfil
  \begin{feynman}{4.5cm}
    \diagram{15}{
      initstate
      -5 0.5 translate
      /init_rand 1117 def
      /n_rand 30 def
      /lambda 360 def
      /fluct_ampl 0.0 def
      /rand_ampl 0.0 def
      /field_line 
      {
	/fluct exch def
	/lambda rand int_max div 180 mul 180 add def
	/init_phase rand int_max div 360 mul def
	newpath
	rand int_max div 1.5 mul 0.25 add dup
	1.5 mul
        rand int_max div 4.5 mul 0.25 add add
	2 copy
	add 10 div 0.9 mul 1 exch sub 0 0 setrgbcolor
	moveto
	/dx 0.02 def
	0 dx 5 
	{
	  lambda mul init_phase add sin fluct mul dx mul 
	  1 rand int_max div rand_ampl mul 2 mul rand_ampl sub add mul
	  dx 
	  exch
	  rlineto
	} for
	stroke
      } def
      gsave
      0 0 translate
      initclip
      newpath 
      0 0 moveto
      0 5 rlineto
      2 3 rlineto
      0 -5 rlineto
      closepath gsave clip newpath
      << /ShadingType 2
         /ColorSpace /DeviceRGB
         /Coords [2 0 0 0]
         /AntiAlias true 
         /Function << /FunctionType 2
                 /Domain [0 1]
                 /C0 [0 0 0]
                 /C1 [0.8 0.8 0.8]
                 /N 0.6
              >>
      >> shfill
      grestore initclip
      0.06 setlinewidth
      fgcolor
      0 0 moveto
      0 5 rlineto
      2 3 rlineto
      0 -5 rlineto
      -2 -3 rlineto stroke
      gsave
      colorb
      0.06 setlinewidth
      init_rand srand
      n_rand {fluct_ampl field_line} repeat
      gsave
      5 0 translate
      initclip
      newpath 
      0 0 moveto
      0 5 rlineto
      2 3 rlineto
      0 -5 rlineto
      closepath gsave clip newpath
      << /ShadingType 2
         /ColorSpace /DeviceRGB
         /Coords [2 0 0 0]
         /AntiAlias true 
         /Function << /FunctionType 2
                 /Domain [0 1]
                 /C0 [0 0 0]
                 /C1 [0.9 0.8 0.7]
                 /N 0.6
              >>
      >> shfill
      grestore initclip
      grestore
      colorb
      0.01 setlinewidth
      init_rand srand
      n_rand {fluct_ampl field_line} repeat
      gsave
      5 0 translate
      0.06 setlinewidth
      fgcolor
      0 0 moveto
      0 5 rlineto
      2 3 rlineto
      0 -5 rlineto
      -2 -3 rlineto stroke
      fgcolor
      0.08 setlinewidth
      4 4 moveto 4 0 rlineto currentpoint stroke 0 arrow
      4 4 moveto 0 4 rlineto currentpoint stroke 90 arrow
      4 4 moveto -3 -4.5 rlineto currentpoint stroke -120 arrow
      0.6 tbi
      4.3 7.8 moveto (x) show
      1.4 -0.6 moveto (y) show
      0.6 symb
      8.25 3.8 moveto (h) show
      grestore
    }
  \end{feynman}
  \hfil}

\setbox2\hbox to 5cm{\hfil
  \begin{feynman}{3cm}
    \diagram{15}{
      initstate
      -2 0.5 translate
      /init_rand 1117 def
      /n_rand 30 def
      /lambda 360 def
      /fluct_ampl 1 def
      /rand_ampl 2.5 def
      /field_line 
      {
	/fluct exch def
	/lambda rand int_max div 180 mul 180 add def
	/init_phase rand int_max div 360 mul def
	newpath
	rand int_max div 1.5 mul 0.25 add dup
	1.5 mul
        rand int_max div 4.5 mul 0.25 add add
	2 copy
	add 10 div 0.9 mul 1 exch sub 0 0 setrgbcolor
	moveto
	/dx 0.02 def
	0 dx 5 
	{
	  lambda mul init_phase add sin fluct mul dx mul 
	  1 rand int_max div rand_ampl mul 2 mul rand_ampl sub add mul
	  dx 
	  exch
	  rlineto
	} for
	stroke
      } def
      gsave
      0 0 translate
      initclip
      newpath 
      0 0 moveto
      0 5 rlineto
      2 3 rlineto
      0 -5 rlineto
      closepath gsave clip newpath
      << /ShadingType 2
         /ColorSpace /DeviceRGB
         /Coords [2 0 0 0]
         /AntiAlias true 
         /Function << /FunctionType 2
                 /Domain [0 1]
                 /C0 [0 0 0]
                 /C1 [0.8 0.8 0.8]
                 /N 0.6
              >>
      >> shfill
      grestore initclip
      0.06 setlinewidth
      fgcolor
      0 0 moveto
      0 5 rlineto
      2 3 rlineto
      0 -5 rlineto
      -2 -3 rlineto stroke
      gsave
      colorb
      0.06 setlinewidth
      init_rand srand
      n_rand {fluct_ampl field_line} repeat
      gsave
      5 0 translate
      initclip
      newpath 
      0 0 moveto
      0 5 rlineto
      2 3 rlineto
      0 -5 rlineto
      closepath gsave clip newpath
      << /ShadingType 2
         /ColorSpace /DeviceRGB
         /Coords [2 0 0 0]
         /AntiAlias true 
         /Function << /FunctionType 2
                 /Domain [0 1]
                 /C0 [0 0 0]
                 /C1 [0.9 0.8 0.7]
                 /N 0.6
              >>
      >> shfill
      grestore initclip
      grestore
      colorb
      0.01 setlinewidth
      init_rand srand
      n_rand {fluct_ampl field_line} repeat
      gsave
      5 0 translate
      0.06 setlinewidth
      fgcolor
      0 0 moveto
      0 5 rlineto
      2 3 rlineto
      0 -5 rlineto
      -2 -3 rlineto stroke
      grestore
    }
  \end{feynman}
  \hfil}
\begin{figure}[htbp]
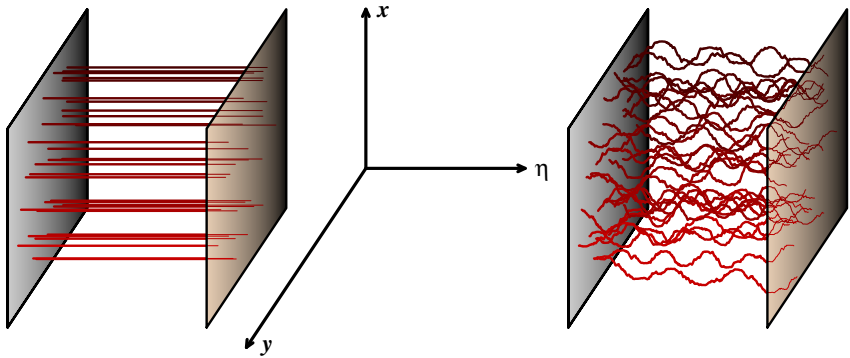

\centerline{
\hfill
\resizebox*{5cm}{!}{\box1}
\hfill
\resizebox*{5cm}{!}{\box2}
\hfill
}
\vskip 2mm
\caption{\label{fig:insta}Left: field lines at leading order. Right:
effect of the instability. (The horizontal axis represents the
space-time rapidity, and the vertical planes the Lorentz contracted
nuclei.)}
\end{figure}
These studies of the initial conditions in heavy ion collisions have
also led to an interesting development, which is expected to have
connections with the issue of thermalization. It has been noted that
the classical color field configurations encountered in the CGC
formalism at leading order are unstable: small rapidity dependent
perturbations grow exponentially in time and eventually become as
large as the classical field itself \cite{insta}. It turns out that
perturbations of this kind are generated by quantum fluctuations
\cite{insta1}, and therefore occur naturally among the NLO
corrections. These quantum fluctuations, amplified by the instability,
would lead to extremely disordered configurations of strong fields
(see figure \ref{fig:insta}). Particles moving in such a medium would
also have a very short mean free path, and the viscosity would be
close to the uncertainty lower bound \cite{ano-eta}. This opens up an
interesting alternative to the strongly coupled scenarios: the
quasi-perfect fluidity inferred from RHIC data may be due to strong
disordered color fields, even if the coupling is rather weak.

\end{document}